\documentclass[journal]{IEEEtran}

\usepackage{graphicx}
\usepackage{amsmath,amssymb}
\usepackage{booktabs}
\usepackage{cite}
\usepackage{hyperref}
\usepackage{amssymb}
\usepackage{orcidlink}  
\usepackage{url}
\usepackage{comment} 
\usepackage{balance}
\usepackage{color}

\begin{document}

\title{A lightweight Outlier Detection for Characterizing Radio- and Environment-Specific  Link Quality Fluctuation in Low-Power Wireless Networks}

\author{Zegeye Mekasha Kidane~\orcidlink{0009-0009-5141-8160}, Waltenegus Dargie~\orcidlink{0000-0002-7911-8081}, \IEEEmembership{Senior Member, IEEE}  
    \thanks{Manuscript submitted on March 13 2026.}
    \thanks{Z. M. Kidane is an external doctoral candidate at the Faculty of Computer Science, Technische Universit{\"a}t Dresden, Germany (zegeye\_mekasha.kidane@mailbox.tu-dresden.de)}
    \thanks{ W. Dargie is with the Faculty of Computer Science, Technische Universit{\"a}t Dresden, 01062 Dresden, Germany (waltenegus.dargie@tu-dresden.de)}
}

\maketitle

\begin{abstract}
The performance of low-power wireless sensing networks can be influenced by both external environmental factors and internal imperfections which often arise due to manufacturing tolerance during mass production. Understanding the conditions and extent of these influences is important not only to achieve high performance and high energy efficiency, but also to carry our environment and radio specific configurations. In this paper we demonstrate, through extensive practical deployments and experiments, the extent to which external and internal factors affect the link quality of low-power wireless sensor networks. Moreover, we propose a lightweight statistical outlier detection technique and define all the parameter thereof in terms of the statistics of both the raw and the predicted link quality metrics (RSSI). Our study considers more than 15 different physical environments consisting of rivers, lakes, bridges, forests, and gardens, as well as four widely employed heterogeneous low-power radios. The results of the experiments suggest that deployment environments, more than hardware imperfections, significantly influence link quality fluctuations, although hardware imperfection does produce a notable imprint on link quality variation.
\end{abstract}

\begin{IEEEkeywords}
Deployment, Internet of Things (IoT), link quality fluctuation, low-power radios, Received Signal Strength Indicator (RSSI), statistical outlier detection,  wireless networks 
\end{IEEEkeywords}

\section{Introduction}

The wide application of Internet of Things (IoT) in outdoor and semi-outdoor environments has increased the demand for resilient low-power wireless communication \cite{chen2023lopdm, chaudhary2022underwater}. Applications such as environmental monitoring, smart agriculture, and structural health monitoring rely on resource-constrained sensing devices that must operate in highly dynamic and harsh propagation conditions \cite{rani2022optimized}. In these scenarios, the radios should be fast to  predict and react to link quality fluctuations \cite{kim2022mobirpl}. 

The Received Signal Strength Indicator (RSSI) is one of the most widely used metrics for analyzing, estimating, and modeling link quality as well as for adaptive transmission power control due to its minimal computational and hardware overhead \cite{dargie2025prediction, tian2025environmental, srinivasan2010empirical}. Nevertheless, RSSI in low-power radios is highly sensitive to environmental and internal factors. Outdoor propagation is strongly affected by factors such as vegetation density, terrain irregularity, and water motion, all of which introduce multipath propagation, shadowing, and time-varying interference. These factors not only lead to significant RSSI fluctuations, but also to poorly correlated or even uncorrelated link quality variations. Deployment- and measurement-based studies in harsh industrial and outdoor environments confirm to this condition, indicating that RSSI instability is the dominant cause of poor link quality characterization and a considerable amount of packet loss in low-power wireless networks \cite{Prasayasith2023}. Device imperfections and manufacturing tolerance in mass production of low-power radios further exacerbate link quality instability and the difficulty to predict and adapt to link quality fluctuations \cite{letafati2020three, liu2021machine}. 

To mitigate these concerns, recent research has focused on anomaly and outlier detection, exploring both statistical methods and machine-learning. The former have the advantage of being computationally affordable and relatively easy to interpret, while the latter offers improved detection accuracy in highly non-stationary environments \cite{Yan7723688}. Surveys and comparative studies emphasize that anomaly detection is becoming a critical component of reliable IoT system modeling, deployment and configuration, particularly for outdoor and large-scale deployments \cite{Adhikari2024,Munoz2024}. However, most existing studies rely on statistics established with the help of homogeneous radio platforms and controlled deployment environments (most of which are lab based). In particular, comparative experimental analyses that jointly consider radio-specific characteristics, environmental variability, and statistical outlier behavior are limited.

This paper aims to fill this gap through a comprehensive experimental and measurement-based study of RSSI fluctuations in low-power wireless networks and statistical outlier detection across heterogeneous IoT radio platforms deployed in multiple, highly dynamic outdoor environments. Four different low-power radio platforms---CC1200, CC2538, nRF52840, and BLE---and five different outdoor scenarios---bridges, forests, lakes, rivers, and gardens---are included in our study. The paper makes the following contributions: (1) Based on extensive experiments and data collection under varied operation conditions and configurations, the paper offers practical and useful insight into the sensitivity of link quality fluctuation in low-power wireless networks. (2) The lightweight statistical outlier detection strategy we propose and employed enables to deal with this sensitivity, isolating correlated link quality fluctuation from uncorrelated (erratic) fluctuation. (3) Furthermore, the sensitivity coefficient we define in the outlier model enables to separately evaluate environmental and radio-specific causes to outliers. 

Overall, the paper addresses {\em the estimation, detection (outliers), and characterization} aspects of low-power wireless links in statistical terms. We argue that the contributions we stated above enable the deployment of resilient networks as well as environment- and radio-specific trade-offs between performance and power consumption.

The remaining part of this work is organized as follows. In Section \ref{sec:relate}, we review related work. In Section \ref{sec:depl}, we describe node deployment. In Section~\ref{sec:link}
 we briefly discuss link quality fluctuation and models for low-power wireless deployments. In Section \ref{sec:models}, we model a wireless link as a random variable and, outliers, as a deviation in the random variable from its mean. The deviation thresholds are expressed in terms of the variance of the random variable as well as a sensitivity coefficient. In Section \ref{sec:result}, we present results based on the statistical evaluation of a large volume of data we gathered from various physical environments using low-power wireless sensor networks. Finally, in Section \ref{sec:conclude}, we give concluding remarks and outline future work.

\section{Related Work}
\label{sec:relate}

Statistical and machine learning–based outlier detection for RSSI and sensor data has been widely studied across wireless sensor networks (WSNs) and IoT systems. Early statistical approaches, such as Z-score or deviation thresholding, focus on lightweight signal filtering that can operate within the computational limits of sensor nodes \cite{Chen2022,Yaro2024}. These methods are often well suited for real-time outlier removal and computational efficiency but their usefulness is limited when the data they deal with exhibit high variability.

Recent studies have sought to combine statistical filtering with machine learning to improve anomaly detection accuracy and robustness. For example, Yan et al. propose a hybrid outlier detection to process health care data. The proposed approach combines a density-based outlier detection with clustering using K-Nearest Neighbor algorithm to balance computational efficiency with detection performance \cite{Yan7723688}. Other hybrid approaches integrate deep reinforcement learning, supervised classifiers (e.g., random forest), and unsupervised anomaly detectors to further enhance detection in heterogeneous IoT environments \cite{SemanticIoT2025}.

Comparative analyses of IoT anomaly detection methods examine how supervised and unsupervised learning models---such as random forests, autoencoders, and support vector machines---can be employed for diverse IoT datasets, highlighting the need for robust preprocessing and scalable architectures \cite{ComNet6G2025}. Hybrid deep learning models combining  convolutional and recurrent networks with autoencoders have also been proposed for spatiotemporal-  and reconstruction-based anomaly detection in smart environment and industrial IoT applications \cite{DengLi2025}.

Survey articles on anomaly detection in IoT and sensor networks further emphasize the rapid growth of AI-based approaches, including deep learning, reinforcement learning, and semantic context–aware methods, which are increasingly adopted to capture complex temporal and spatial relationships in sensor data processing \cite{SemanticIoT2025}. These surveys also discuss the challenges of balancing detection accuracy, computational cost, and resource constraints inherent to IoT nodes \cite{Li2024,ComNet6G2025}.

Similarly, recent research attempts to deal with harsh low-power wireless links by developing advanced time-series anomaly detection techniques that could be relevant for RSSI analysis. Transformer-based architectures for multivariate time series anomaly localization and reconstruction-based models improve detection and localization performance \cite{chen2021learning}. Representation-based and contrastive learning approaches further enhance unsupervised time series anomaly detection by modeling temporal and topological patterns \cite{Vilhes2025PatchTrAD}. Despite the broad literature on anomaly detection in this context, most existing work either focuses on general data or network traffic anomalies or is evaluated on controlled datasets rather than real outdoor deployments. Fewer studies specifically address RSSI time series dynamics across heterogeneous radio platforms under diverse outdoor environmental conditions, leaving a gap in systematic comparative analyses across hardware types and real-world signal variability.

In a previous work \cite{dargie2025prediction}, we propose a lightweight n-step predictor to model and predict received power in low-power wireless networks using RSSI. One of the sensor platforms (Zolertia) and two of the radios  (CC2538 and CC1200) employed in the present work are similar to the ones employed in the previous work. The present work builds on the past experience, but it is markedly different both in purpose and in the deployment environments. In the previous work, outliers were not the focus of our investigation, whereas here they are the primary focus. In the former work, the deployments took place on different water bodies in South Florida, USA, whereas here, they took place in Germany, and consisted of fast-flowing rivers (the Rhine and Elbe Rivers), lakes, forests, bridges, and residential areas. As such, the present work relies entirely on data collected from the deployment environments in Germany. Furthermore, the present work includes two additional low-power radios based on the Adafruit nRF52840 Featherboard architecture.
 

\section{Deployment}
\label{sec:depl}

For our experiments, we employed two widely used sensor node architectures: Zolertia RevB\footnote{\url{https://doc.riot-os.org/group\__boards\__remote-revb.html}} and Adafruit nRF52840 Feather\footnote{\url{https://www.adafruit.com}}. The former ran the  Contiki\cite{oikonomou2022contiki} operating system, whereas the latter ran the RIOT\cite{baccelli2018riot} operating system.  

The Zolertia RevB architecture is based on the 32-bit ARM Cortex-M3 processor with a 32 MHz clock speed and integrates two IEEE 802.15.4-compliant low-power radio modules from Texas Instruments: a sub-gigahertz radio module (CC1200~\cite{ti_cc1200}) with 50 kbit/s and a 2.4 GHz radio module (CC2538~\cite{ti_cc2538}) with 250 kbit/s. The latter has 16 channels, each with a 2 MHz bandwidth. The channels are separated by a 5 MHz wide guard band. The radio module has a range of approximately 100 m, a sensitivity of -97 dBm, and an adjustable transmit power with a peak value of 7 dBm (the RSSI value is therefore between -97 dBm and +7 dBm). The CC1200 has a wide operating frequency range (137–158.3, 164–190, 205–237.5, 274–316.6, 410–475, 820–950 MHz) and a nominal range of approximately 4 km. Its sensitivity depends on the transmission rate (–123 dBm at 1.2 kbit/s and –109 dBm at 50 kbit/s); its maximum transmit power is 16 dBm (therefore, the radio's RSSI value is between –123 dBm and 16 dBm). For our experiments, the transmit power of both radios was set to 0 dBm. The CC2538 radio was configured to transmit 25 packets per second. The CC1200 radio was configured to transmit 3 packets per second at 868.5 MHz.

The Adafruit nRF52840 Feather Node~\cite{nordic_nrf52840} is an energy-efficient and high-performance system-on-a-chip (SoC) with a 32-bit ARM Cortex-M4F processor clocked at 64 MHz. It integrates a multi-protocol SoC radio stack that can be configured as Bluetooth Low Energy (BLE), Thread, or ZigBee. The radio module's transmit power is adjustable to 0 dBm, 4 dBm, and 8 dBm, with an RSSI value between -97 dBm and +8 dBm. For our experiments, the transmit power was set to 0 dBm. In IEEE 802.15.4 mode, 10 packets per second were transmitted, and in BLE mode, 5 packets per second.

The RSSI value is a measure of the received signal strength in dBm. According to the IEEE 802.15.4 specification, it is measured across the entire channel bandwidth (2 MHz for CC2538 and nRF52840, including BLE; 12.5 kHz for CC1200) and estimated over eight symbol periods (128 µs). It is read directly from an 8-bit, signed 2’s complement register. In the CC1200 radio, the RSSI is sampled at a resolution of  0.5 dBm; in the others, at a resolution of 1 dBm.



\begin{table}[h!]
    \caption{Node Configuration.}
    \label{tab:node_specs}
    \centering
    \resizebox{\columnwidth}{!}{%
    \begin{tabular}{l l l c c  l}
        \hline
        \textbf{Node} & \textbf{Radio} & \textbf{Frequency} & \textbf{Pkt/s} &  \textbf{Data Rate (kbps)} & \textbf{OS} \\
        \hline
        Zolertia RevB & CC2538 & 2.4 GHz & 25  & 250 & Contiki \\
        Zolertia RevB & CC1200 & Sub-1 GHz & 3  & 1.2 & Contiki \\
        nRF52840 Feather & nRF52840 & 2.4 GHz & 10 & 250 & RIOT \\
        nRF52840 Feather & BLE & 2.4 GHz & 5  & 125 & RIOT \\
        \hline
    \end{tabular}%
    }
\end{table}

\begin{figure*}[h!]
    \centering
        \includegraphics[width=0.23\textwidth]{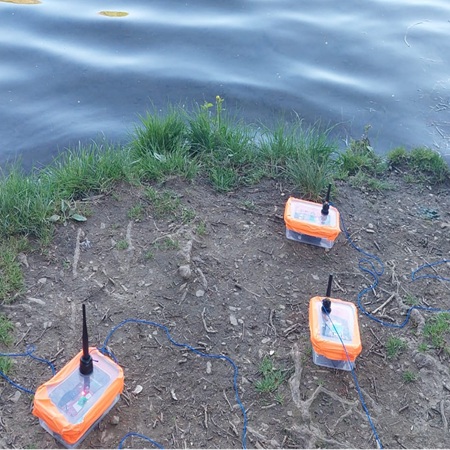}        
        \includegraphics[width=0.23\textwidth]{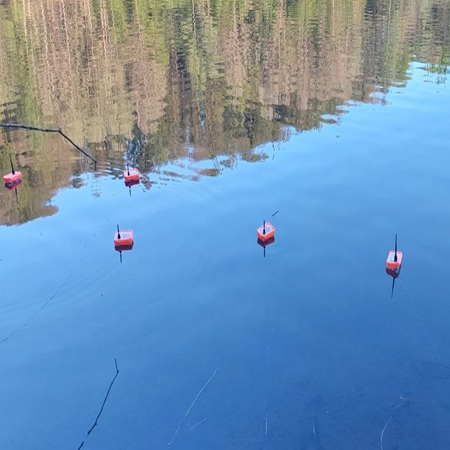}
        \includegraphics[width=0.23\textwidth]{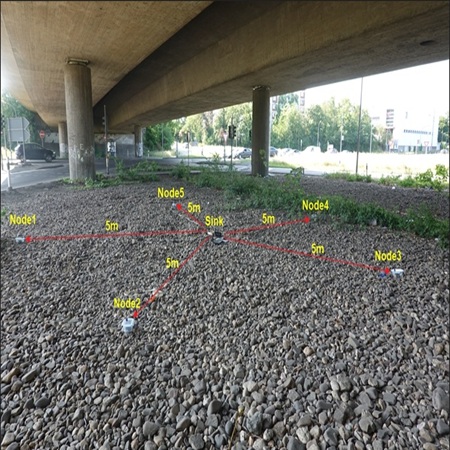}
        \includegraphics[width=0.23\textwidth]{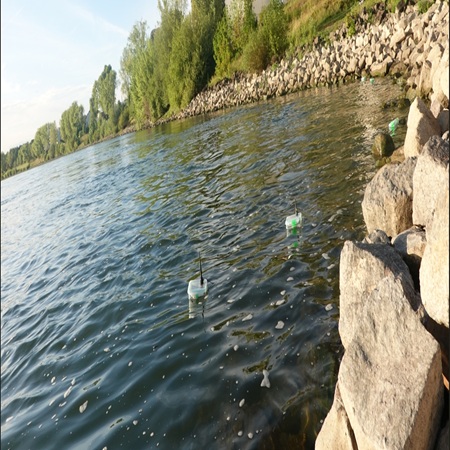}
      \caption{Prototype deployment in West Germany. From left to right:  Sensor nodes in waterproof boxes before deployment. Deployment on a small lake (LK); deployment under Cologne Bridge (BG), along the bridge structure; and deployment on the Rhine River (RV).}
    \label{fig:deployment}
\end{figure*}

Our sensor networks were deployed in 15 different physical environments. Some of the nodes were placed inside waterproof boxes to protect them from extreme weather, rain, and/or surface water. A network of 6 Zolertia nodes (with CC2538 and CC1200 radios) was deployed in West Germany in five different places (on the Rhine River (RV), a small lake (LK), under a bridge (BG), on a forest (FR), and in a residential garden (GG)). The deployment environments were chosen because they represent various real-world environments. Fig.~ \ref{fig:deployment} displays the preparation and deployments of the sensor networks. The network had a star topology in which five leaf nodes transmitted packets to a sixth central node, which played the role of a sink. The distance between the nodes varied between 5 to 10 m, depending on the specific deployment environment and surrounding objects; the distance on the river and the lake was affected by water motion. Since, we were interested in modeling outliers in individual wireless links, the difference in transmission distance was of secondary concern. During the entire experiments, we made sure that no link was permanently broken. In all the experiments, the antennas were vertically oriented (the nodes employed Herdio Marine waterproof Antennas which were approximately 16 cm high above ground or water surface). Upon receiving a packet from its leaf nodes, the sink extracted the RSSI and store the data along with a timestamp. For the deployments in the western part of Germany, a single experiment lasted about 20 minutes. For each deployment and configuration, we conducted at least five independent experiments on different days under different weather conditions. 

Likewise, the Adafruit nRF52840 Feather nodes were deployed in 10 different places in the eastern part of Germany (the city of Dresden). Two different groups of students carried out the deployments, each group targeting 5 deployment environments: the Elbe river, a busy bridge, a forest, a small lake, and a big public park. For these deployments, a network consisted of 3 nodes forming a mesh topology. Here as well, the distance between the sensor nodes varied from 5 to 10 meter, depending on the nature of the deployment environments. Each node broadcast packets to the other nodes in the network, and each node locally kept a record of the packets it received along with the ID of the transmitter, timestamps, and the RSSI of the received packets. For the deployments in Dresden, the duration of a single experiment was 30 minutes. Table~\ref{tab:nodes} summarizes the deployment in both parts of Germany.

\begin{table}[!h]
\centering
\caption{Sensor Node Deployment Across Radio Platforms}
\label{tab:nodes}
\begin{tabular}{lc}
\toprule
\textbf{Radio Platform} & \textbf{Number of Nodes} \\
\midrule
CC1200     & 5 \\
CC2538    & 5 \\
nRF52840  & 3 \\
BLE       & 3 \\
\bottomrule
\end{tabular}
\end{table}


\section{Wireless Link Quality}
\label{sec:link}

In low-power wireless sensor networks, the received signal strength indicator (RSSI) is commonly used as a physical-layer metric for link quality assessment and abnormal behavior detection. The different radios we employed for our study, despite differences in operating frequency and modulation, share similar RSSI measurement limitations due to low-cost and energy-efficient hardware design. In general, the received signal strength (in dBm) at time $t$ can be decomposed into four components:
\begin{equation}
\mathbf{R}_t = P_t - PL(d) + D_t + \eta_t
\label{equ:rssi}
\end{equation}
All the terms on the right hand side are expressed in terms of dBm as well. $P_t$ denotes the transmit power; $PL(d)$ represents large-scale path loss as a function of the distance separating the transmitter and the receiver; $E_t$ captures deployment or environment-induced variations, and $\eta_t$ represents stochastic noise and fast fading effects.

\subsubsection{Transmit Power}

The transmit power $P_t$ is constrained and typically selected from a discrete set of hardware-supported levels (ref. to Section~\ref{sec:depl}). 

\subsubsection{Large-Scale Path Loss}

The large-scale path loss term $PL(d)$ models the deterministic attenuation of the wireless signal with distance and is commonly expressed using the log-distance path loss model:
\begin{equation}
PL(d) = PL(d_0) + 10n \log_{10}\left(\frac{d}{d_0}\right)
\end{equation}
where $d_0$ denotes a reference distance and $n$ is the path loss exponent. Radios operating at 2.4~GHz, including the CC2538 and nRF52840 in both IEEE~802.15.4 and BLE modes, experience higher free-space attenuation and increased sensitivity to obstructions and human body shadowing. In contrast, the sub-GHz CC1200 exhibits lower path loss and improved penetration, resulting in more stable RSSI behavior over distance~\cite{ti_cc1200}.

\subsubsection{Environment-Induced Variations}

The term $D_t$ accounts for slow-varying environment-induced effects such as shadowing, reflection, scattering, and temporal variations caused by moving objects or human activity. These effects are particularly relevant in indoor and semi-static deployment scenarios. In BLE, frequency hopping across multiple channels introduces additional variability, as individual channels experience distinct multipath conditions~\cite{nordic_nrf52840}. IEEE~802.15.4 links typically operate at a fixed channel, leading to RSSI variations that are more strongly influenced by local multipath characteristics. For the sub-GHz CC1200, the longer carrier wavelength reduces sensitivity to small-scale obstructions, resulting in smoother temporal RSSI variations.

\subsubsection{Noise and Fast Fading}

The term $\eta_t$ represents fast-varying stochastic effects, including thermal noise, receiver front-end noise, fast multipath fading, co-channel interference, and system (manufacturing) imperfections or tolerance. For the CC2538, RSSI values are averaged over multiple symbols, which reduces instantaneous variance at the cost of increased measurement latency~\cite{ti_cc2538}. The nRF52840 reports RSSI on a per-packet basis in both IEEE~802.15.4 and BLE modes, with BLE measurements being particularly sensitive to short packet durations and channel hopping~\cite{nordic_nrf52840}. The CC1200 provides configurable RSSI filtering options, enabling a trade-off between responsiveness and noise suppression~\cite{ti_cc1200}. Due to these factors, $\eta(t)$ is typically modeled as a zero-mean Gaussian random process with variance $\sigma^2$.

\subsection{Wide-Sense Stationary Property}

\begin{figure*}[t]
\centering
\includegraphics[width=0.8\textwidth]{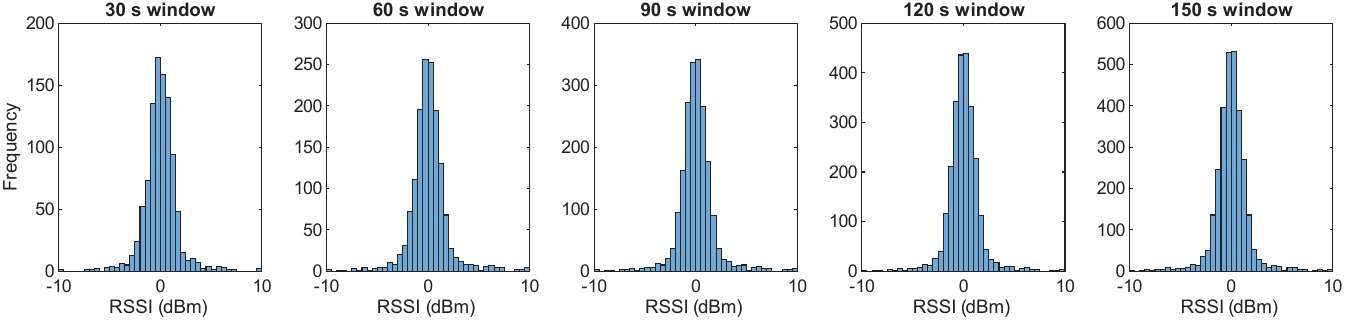}
\caption{Histograms of $\Delta RSSI$ for five different time windows (Radio: CC2538. Location: Rhine River).}
\label{fig:stationary}
\end{figure*}

The condition for a wide-sense stationary stochastic process is that some of the essential statistical features (mean and variance) are invariant to time shifts. This condition could be observed in the low-power networks we deployed for all the radios, provided that the time interval of interest is in the order or several seconds. Fig.~\ref{fig:stationary} shows the histograms of the change in RSSI for the deployment we carried out on the Rhine River using the CC2538 radio (the least reliable, as we shall show). The time windows we considered were 30, 60, 90, 120, and 150 seconds. Within these time windows, the statistics of the change in RSSI exhibited no appreciable difference in the shape of the distribution, mean (0.0095, 0.0206, 0.0166, 0.0139, 0.0045) and variance (3.58, 5.48, 4.57, 3.87, 4.47). In the sections that follow, we assume that the stationary condition applies to the outlier detection approach we propose. It is also implied that the time window which concerns us is shorter than 180 seconds.

\section{Modeling Outliers}
\label{sec:models}

Let $\mathbf{R}_t$ denote the raw RSSI measurement at a discrete time index \(t\), expressed as a random variable. To suppress high-frequency noise while preserving the slowly varying propagation trend, we apply an exponential moving average ($\mathbf{E}_t$), equivalent to a first-order causal infinite impulse response (IIR) low-pass filter. Here, the choice of a low-pass filter is deliberate, as outliers are typically erratic, high-frequency variations in the input time series which do not allow meaningful system adaptations. The filtered RSSI sequence can be expressed as:
\begin{equation}
\mathbf{\mathbf{E}}_t = \alpha \mathbf{R}_t + (1-\alpha)\mathbf{E}_{t-1}, \quad 0 < \alpha \leq 1
\label{equ:ema}
\end{equation}
with initialization $\mathbf{E}_0 = R_0$. If we use the Kalman nomenclature \cite{kalman1960new}, the first term of the RHS of Equation~\ref{equ:ema} corresponds to a fresh measurement for time $t$, and the second term corresponds to a prediction for the same time. From a signal processing perspective, the EMA implements a single-pole low-pass filter whose effective bandwidth is controlled by the smoothing factor \(\alpha\). Smaller values of \(\alpha\) correspond to stronger attenuation of fast fading and measurement noise, while larger values increase responsiveness to short-term RSSI fluctuations. The EMA output thus serves as an estimate of the underlying mean RSSI component associated with large-scale propagation effects. 


\begin{table}[h!]
\centering
\caption{Notation used throughout EMA smoothing and RSSI outlier detection.}
\label{tab:notation}
\begin{tabular}{l p{4cm} c} 
\toprule
\textbf{Symbol} & \textbf{Meaning} & \textbf{Units / Notes} \\
\midrule
\(\mathbf{R}_t\) & Raw RSSI measurement at time \(t\) & dBm \\
\(\mathbf{E}_t\) & Exponential moving average of RSSI at time \(t\) & dBm \\
\(\mathbf{z}_t\) & abs. diff between \(\mathbf{R}_t\) and  \(\mathbf{E}_{t-1}\) & dBm \\
\(\alpha\) & EMA smoothing factor & dimensionless, \\ 
\(\sigma_R^2\) & Variance of \(\mathbf{R}_t\) & (dBm)$^2$ \\
\(\sigma_E^2\) & Variance of \(\mathbf{E}_t\) & (dBm)$^2$ \\
\(\sigma_z^2\) & Variance of \(\mathbf{z}_t\) & (dBm)$^2$ \\
\(k\) & Sensitivity parameter for outlier detection & dimensionless \\
\(t\) & Discrete time index & samples \\
\(\mathbf{1}(\cdot)\) & Indicator function: 1 if condition true, 0 otherwise & dimensionless \\
\bottomrule
\end{tabular}
\end{table}

\subsection{Variance-based selection of the EMA parameter}
\label{subsec:alpha_variance}

Under stationary and steady-state conditions, the variance of $E_{t-1}$ is a function of the variances of the past (up to $t-1$) RSSI values. Following the Central Limit Theorem \cite{papoulis2002probability}, its distribution is more likely normal and its correlation with $R_t$ is not as strong as $R_t$'s correlation with $R_{t-1}$. It can be easily shown that the variance of $E_t$ can be expressed as: 
\begin{equation}
\label{eq:ema_variance}
\begin{aligned}
\operatorname{Var}(\mathrm{E}_t)
& = \alpha^2 \operatorname{Var}(\mathbf{R}_t)
 + (1-\alpha)^2 \operatorname{Var}(\mathrm{E}_{t-1}) \\
&\quad + 2\alpha(1-\alpha)\operatorname{Cov}(\mathbf{R}_t,\mathrm{E}_{t-1})
\end{aligned}
\end{equation}
Since we assume that \(\mathbf{R}_t\) is weakly correlated with \(\mathrm{\mathbf{E}}_{t-1}\), the covariance term in Equation~\ref{eq:ema_variance} can be neglected. Moreover, under the steady state assumption:
\begin{equation}
\operatorname{Var}(\mathrm{\mathbf{E}}_t)
\approx \operatorname{Var}(\mathrm{\mathbf{E}}_{t-1})
\approx \operatorname{Var}(\mathrm{\mathbf{E}})
\end{equation}
Consequently,
\begin{equation}
\mathrm{Var}(\mathbf{E}) = \frac{\alpha}{2-\alpha}\,\mathrm{Var}(R)
\label{equ:ema_variance_steady}
\end{equation}
If we let \(\sigma_R^2 = \mathrm{Var}(R)\) and \(\sigma_{\mathbf{E}}^2 = \mathrm{Var}(\mathbf{E})\), Solving Equation \eqref{equ:ema_variance_steady} for \(\alpha\) yields:
\begin{equation}
\alpha = \frac{2\sigma_{\mathbf{E}}^2}{\sigma_R^2 + \sigma_{\mathbf{E}}^2}
\label{equ:alpha_est}
\end{equation}

Equation~\eqref{equ:alpha_est} shows that the EMA smoothing parameter is directly determined by the ratio of the filtered and unfiltered RSSI variances. A small variance reduction (\(\sigma_{\mathbf{E}}^2 \approx \sigma_R^2\)) results in \(\alpha \rightarrow 1\), corresponding to minimal smoothing and high responsiveness. Conversely, a strong variance reduction (\(\sigma_{\mathbf{E}}^2 \ll \sigma_R^2\)) yields \(\alpha \rightarrow 0\), indicating aggressive low-pass filtering. This variance-based formulation provides an intuitive and principled mechanism for tuning the EMA.


\subsection{Statistical Outlier Detection}

Chebyshev's Inequality \cite{papoulis2002probability} serves as the basis for our outlier detection:
\begin{equation}
    \label{eq:cheb}
    P \left \{ | \mathbf{R}_t - \eta_R | \geq \epsilon \right \} \leq \frac{\sigma^2_R}{\epsilon^2}
\end{equation}
where $\mathbf{R}_t$ represents link quality fluctuation at time $t$ modeled as a random variable and $\eta_R$ and $\sigma^2_R$ are its mean and variance, respectively. For a continuous random variable, the probability that the values of the random variable lying outside of the interval specified by the left side term in Equation~\ref{eq:cheb} is given by:
\begin{align}
    \label{eq:cheb1}
    P \left \{ | \mathbf{R}_t - \eta_R | \geq k \right \} & = \int_{-\infty}^{\eta_R - \epsilon } f(R) \; dR + \int_{\eta_R + \epsilon }^{\infty} f(R) \; dR \nonumber \\
    & = \int_{|R-\eta_R| \geq \epsilon } f(R) \; dR 
\end{align}
where $f(R)$ is the probability density function of $\mathbf{R}_t$ and $R$ is the specific value $\mathbf{R}_t$ can take, with a probability associated with $f(R)$. The  variance of the random variable is by definition:
\begin{equation}
    \label{eq:cheb3}
    \sigma^2_R = \int_{-\infty}^{\infty} (R - \eta_R )^2 f(R) \; dR  
\end{equation}
Since all the terms in the integral of Equation~\ref{eq:cheb3} are positive and the integral goes from $-\infty$ to $\infty$, the following inequality holds:
\begin{equation}
    \label{eq:cheb4}
    \sigma^2_R \geq \epsilon^2 \int_{|\mathbf{R}_t -\eta_R| \geq \epsilon } f(R) \; dR
\end{equation}
Rearranging the terms in Equation~\ref{eq:cheb4} yields Chebyshev's Inequality, since the integral on the right hand side equals $P\{ |\mathbf{R}_t - \eta_R | \geq k \}$. 

\begin{figure*}
    \centering
    \includegraphics[width=0.8\linewidth]{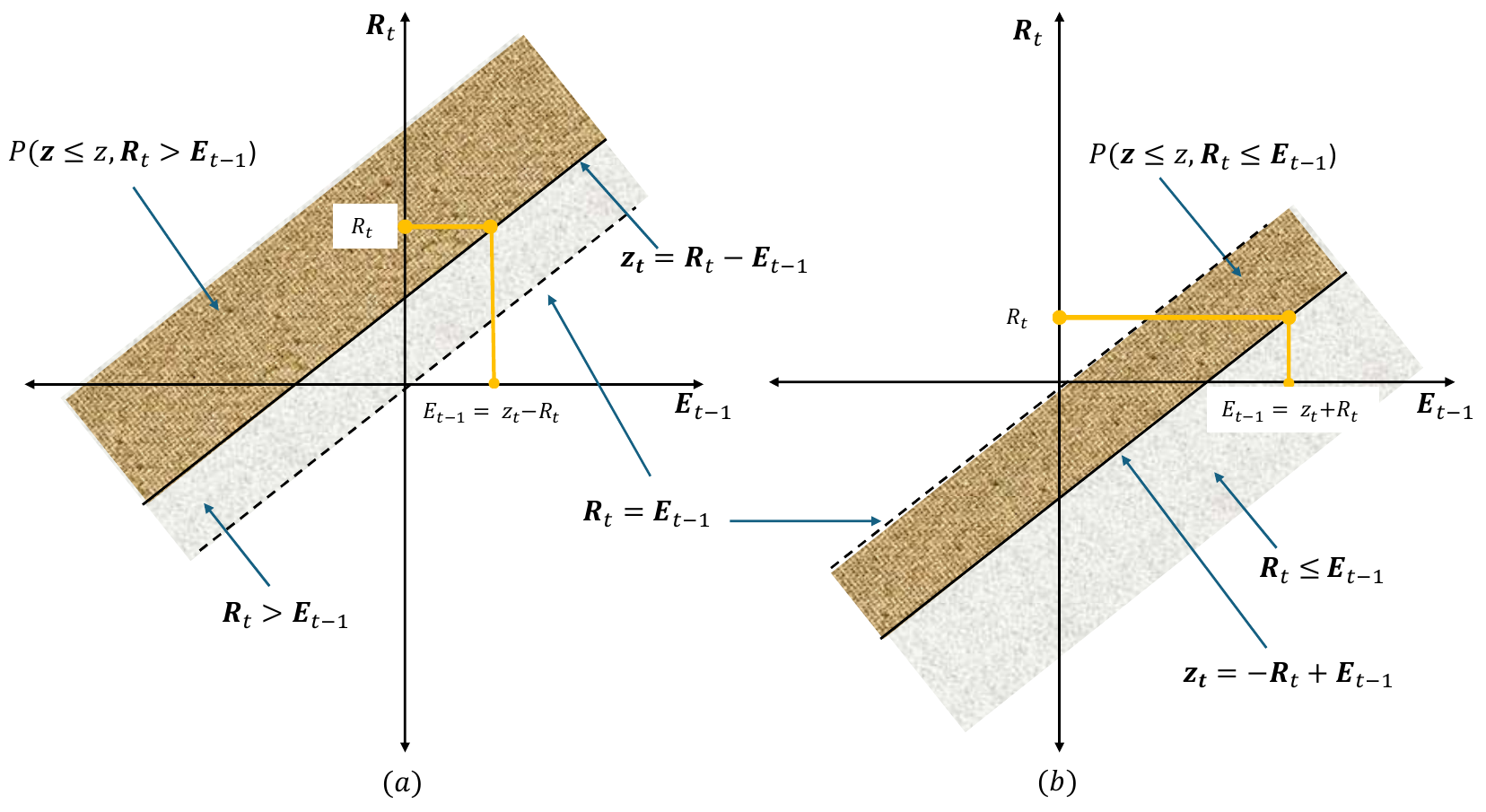}
    \caption{The regions of interest to determine the probability distribution function of $\mathbf{z}_t$}
    \label{fig:z}
\end{figure*}

The inequality sets an upper limit to the probability that the link quality deviates from its mean in terms of a sensitivity coefficient, $k$. Considering the sensitive nature of $\mathbf{R}_t$, applying Chebyshev's Inequality on it directly makes outlier modeling highly unstable. Instead we model outlier as the absolute deviation of $\mathbf{R}_t$ from $\mathbf{E}_{t-1}$. If we let:
\begin{equation}
\label{eq:z}
    \mathbf{z}_t = |\mathbf{R}_t - \mathbf{E}_{t-1}| 
\end{equation}
We can now specify the outlier condition for a given confidence level $\delta = \sigma_z^2/\epsilon^2$, as follows (ref. to Equation~\ref{eq:cheb4}\footnote{Now the random variable of interest is $\mathbf{z}_t \geq 0$, and the condition of outlier is: $P\left \{ \mathbf{z}_t \geq E [\mathbf{z}_t] + \epsilon \right \} = \int_{\eta_z + \epsilon}^{\infty} f(z) \; dz \leq \sigma^2_z /\epsilon^2
$.}):
\begin{equation}
\label{eq:outlier}
\text{Outlier}_t =
\begin{cases}
 1, & \mathbf{z}_t  \geq  \mathbb{E}[\mathbf{z}_t ]  +  \epsilon  \\
 0, & \text{otherwise}
\end{cases}
\end{equation}
Where $\mathbb{E}[.]$ is the expected value of the function.
Since $\mathbf{R}_t$ and $\mathbf{E}_{t-1}$ are both random variables, so is $\mathbf{z}_t$ and it is important to determine its probability density function in order to evaluate Equation~\ref{eq:outlier}. For this task, we have to determine the region of interest in the $(\mathbf{R}_t,\mathbf{E}_{t-1})$ plane. 

\subsection{The Density Function of $\mathbf{z}_t$}

The absolute value term in Equation~\ref{eq:z} is evaluated differently depending on the relative magnitudes of the random variables:
\begin{equation}
\label{eq:z1}
    \mathbf{z}_t = 
    \begin{cases}
         \mathbf{R}_t - \mathbf{E}_{t-1}, & \mathbf{R}_t > \mathbf{E}_{t-1}  \\
         - \mathbf{R}_t + \mathbf{E}_{t-1}, & \mathbf{R}_t \leq \mathbf{E}_{t-1}
    \end{cases}
\end{equation}
In the following, we determine the probability distribution function of $\mathbf{z}_t$ first: $F(z) = P\{\mathbf{z}_t \leq z \}$; once this is accomplished, we differentiate $F(z)$ with respect to $z$ to determine the probability density function $f(z)$. The density function contains all the information we need about $\mathbf{z}_t$.

As can be seen in Figure~\ref{fig:z}, the two cases in Equation~\ref{eq:z1} are separated by the dashed line $\mathbf{R}_t = \mathbf{E}_{t-1}$. When $\mathbf{R}_t > \mathbf{E}_{t-1}$, the region of interest to determine $F(z)$ is depicted in Figure~\ref{fig:z} (a). This is the intersection of two events, namely, $P\{\mathbf{z} \leq z\}$ and $P\{\mathbf{R}_t > \mathbf{E}_{t-1} \}$:
\begin{equation}
    \label{eq:z2}
    F(z) = \int_{-\infty}^{\infty} \int_{-\infty}^{R = z + E} f(R, E) \; dR \; dE
\end{equation}
where $f(R,E)$ is the joint density function. We have dropped the time indices in the integration and will continue to do so as long as the context is clear. For the case where $\mathbf{R}_t \leq \mathbf{E}_{t-1}$, the region of interest is depicted in Figure~\ref{fig:z} (b). Unlike the previous case, the intersection $P\{\mathbf{z} \leq z, \mathbf{R}_t \leq \mathbf{E}_{t-1} \}$ is not straightforward to integrate; instead, it is much easier to deal with its complement: $P\{\mathbf{z} > z, \mathbf{R}_t \leq \mathbf{E}_{t-1} \} $. Hence,
\begin{equation}
    \label{eq:z3}
    F(z) = 1 -  \int_{-\infty}^{\infty} \int_{R = E-z}^{\infty} f(R, E) \; dR \; dE
\end{equation}

In order to get $f(z)$, we apply Leibnitz's differentiation rule\footnote{Given we have a function of two random variables: 
\[F(t) = \int_{a(z)}^{b(z)}f(x, z) \; dx \]
\[ \frac{dF(z)}{dz}  = \frac{db(z)}{dz} f(b(z), x) - \frac{da(z)}{dz} f(a(z), x) + \int_{a(z)}^{b(z)}\frac{d f(x, z)}{dz} \; dx\]
} on Equations~\ref{eq:z2} and ~\ref{eq:z3} to obtain the following results\footnote{For Equation~\ref{eq:z2}: 
\begin{align}
f(z) & = \frac{dF(z)}{dz} = \int_{-\infty}^{\infty} \frac{d}{dz} \int_{\infty}^{z + E} f(R,E) \; dR \; dE \nonumber \\
     & = \int_{-\infty}^{\infty} \left ( 1 \; . f(z+E, E) - 0  \; . + \int_{\infty}^{z + E} \frac{\partial f(R, E)}{\partial z} dR \right ) \; dE  \nonumber \\
     & = \int_{-\infty}^{\infty} f( \underbrace{z+E}_{R}, E) dE
\end{align}
}:
\begin{equation}
    \label{eq:z4}
    f(z)= 
    \begin{cases}
        \int_{-\infty}^{\infty} f(z+E, E) \; dE, & \mathbf{R}_t > \mathbf{E}_{t-1} \\
        \int_{-\infty}^{\infty} f(E-z, E) \; dE, & \mathbf{R}_t \leq \mathbf{E}_{t-1}
    \end{cases}
\end{equation}
If $\mathbf{R}_t$ and $\mathbf{E}_{t-1}$ are (jointly) normal, then $\mathbf{z}$ is also normal. If we regard $\text{Cov(}\mathbf{R}_t, \mathbf{E}_{t-1}\text{)} \approx 0$ (Equation~\ref{eq:ema_variance}), then Equations~\ref{eq:z4} becomes\footnote{
When $R_t$ and $E_{t-1}$ are jointly normal, then their density is of the following type:
\[
f(R_t, E) \sim \text{exp} \left \{ -\frac{1}{2(1-r^2)} \left ( A^2   - 2r AB  + B^2  \right ) \right \}
\]
where,
\[
r = \text{Cov}(R_t, E_{t-1})
\]
\[
A^2 = \frac{(R_{t} - \eta_R)^2}{\sigma_R^2}
\]
\[
B^2 = \frac{(E_{t-1} - \eta_E)^2}{\sigma_E^2}
\]
If $r \approx 0$, then the density function becomes the multiplication of two exponent terms.
}:
\begin{equation}
    \label{eq:z5}
    f(z)= 
    \begin{cases}
        \int_{-\infty}^{\infty} \;  f_R(E+z) f(E) \; dE, & \mathbf{R}_t > \mathbf{E}_{t-1} \\
        \int_{-\infty}^{\infty} \; f_R(E-z)f( E) \; dE, & \mathbf{R}_t \leq \mathbf{E}_{t-1}
    \end{cases}
\end{equation}
where $f_R$ is used to imply the density of $\mathbf{R}_t$. When $f(R)$ and $f(E)$ are normally distributed with the same mean, then $\mathbf{z}_t$ is also normal with zero mean (truncated, of the sort $f(z) \sim e^{-z^2/2\sigma_z^2}, \; z \geq 0$). Its variance is given as:
\begin{equation}
    \label{eq:z6}
    \sigma_z^2 = \sigma_R^2 + \sigma_E^2
\end{equation}

Since we now have the variance of $\mathbf{z}_t$, we can describe the sensitivity of a wireless link (due either to the environment or the radio) in terms of it. To do this, we  modify the outlier condition in Equation~\ref{eq:outlier} as follows:

\begin{equation}
\label{eq:k1}
\mathbf{z}_t \geq k \sigma_z \quad  k \sigma_z = \mathbb{E}[ \mathbf{z}_t ] + \epsilon 
\end{equation}
so that: 
\begin{equation}
    \label{eq:k2}
k = \frac{\mathbb{E} \left [ \mathbf{z}_t \right ] + \epsilon  }{\sigma_z}
\end{equation}
Or, in terms of the confidence level, $\delta$,
\begin{equation}
    \label{eq:k3}
k = \frac{\mathbb{E} \left [ \mathbf{z}_t \right ] }{\sigma_z} + \frac{1}{\sqrt{\delta}} \approx \frac{1}{\sqrt{\delta}}
\end{equation}
As can be seen, $k$ is directly proportional to the variance of $\mathbf{z}_t$ and inversely proportional to the square root of the confidence level. For a fixed confidence level, the larger the variance of $\mathbf{z}_t$, the larger is the sensitivity coefficient, signifying an unstable link. The sensitivity coefficient regulates detection aggressiveness under noisy and time-varying RSSI conditions and primarily influences the rate and stability of detections.

\section{Results}
\label{sec:result}

In this section, we apply the model we propose to analyze the measurements we collected from the various environments using the four low-power radios. Our objective is to assess the impact of environment vs device imperfection on the wireless links the networks established.

\subsection{RSSI Dynamics and Outlier Patterns}

\begin{figure*}[!ht]
\centering
\includegraphics[width=1.0\textwidth]{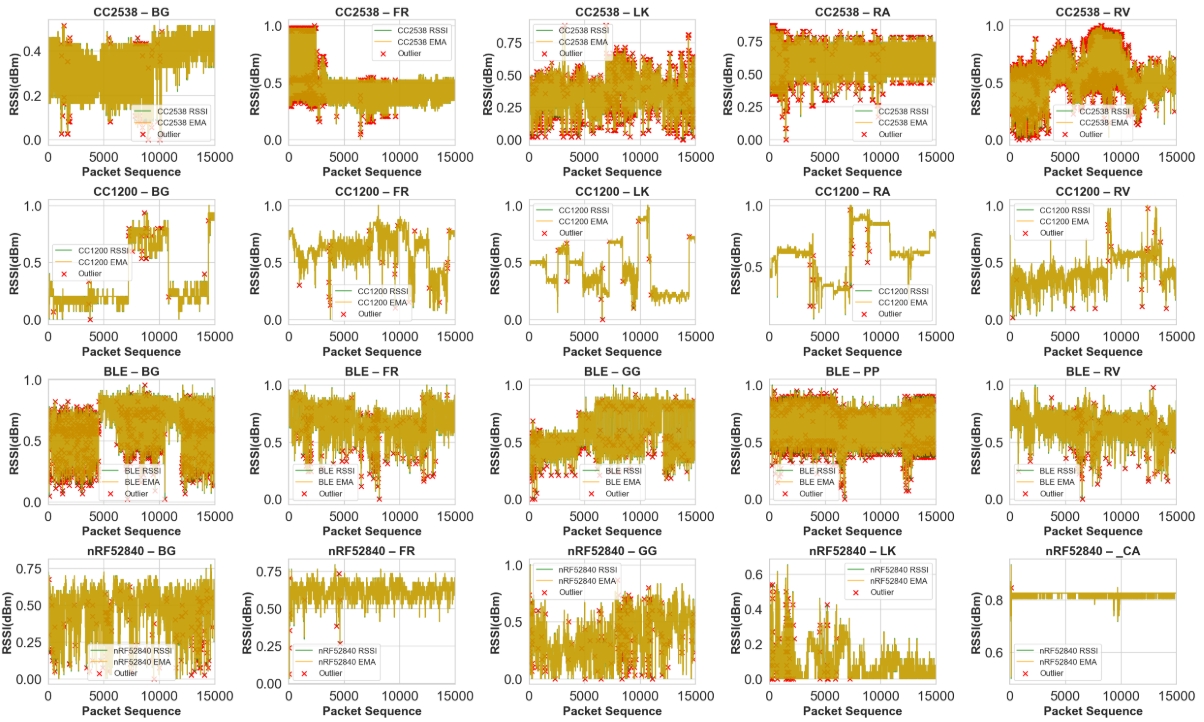}
\caption{Normalized RSSI variations in the 15 deployment environments using the four low-power radios. The red crosses mark the detected RSSI outliers.}
\label{fig:rssi}
\end{figure*}

Figure~\ref{fig:rssi} illustrates the temporal evolution of RSSI across the five groups of environments for all the radio platforms. The observed RSSI dynamics are governed by a combination of large-scale path loss, medium-scale shadowing, and small-scale fading. Environments with high structural complexity, notably bridges and forests, exhibit considerable RSSI fluctuation. These environments introduce persistent multipath propagation and shadowing effects, leading to frequent deviations from the EMA baseline. Statistical analysis using  a one-way ANOVA suggests that the mean RSSI differs significantly across environments for all radios ( $p < 0.05 $), assuming independence of observations.  

In many random-effects experiments, the primary objective is to estimate the overall mean by using an unbiased estimator and constructing a 95\% confidence interval $100(1-\delta)\%$ around it \cite{montgomery2022design}. To determine the 95\% confidence interval (CI) for the overall mean of RSSI variation based on the experimental results, it is necessary to compute the mean squares and determine the corresponding lower (L) and upper (U) confidence bounds, such that $L \leq \mu \leq U$. The results are reported in F-Statistic together with the associated measurement uncertainty, expressed in accordance with the Guide to the Expression of Uncertainty in Measurement (GUM) using a coverage factor of k=2. For example the \textit{BLE} environmental result in Table \ref{tab:f_statistics_comparison}, the reported value of F-statistics is \(149{,}256 \pm 105{,}000\) with \(k=2\) and \(df_{in} = 89{,}876\). The corresponding standard uncertainty is \(105{,}000/2 = 52{,}500\). The lower and upper confidence bounds are obtained as \(L = 149{,}256 - 105{,}000 = 44{,}256\) and \(U = 149{,}256 + 105{,}000 = 254{,}256\). Therefore, the true mean value is expected to satisfy \(44{,}256 \leq \mu \leq 254{,}256\) with approximately \(95\%\) confidence.

A one-way ANOVA was performed to assess RSSI mean across deployment environments and radios. Device-level performance consistency reflects the consistency of RSSI measurements across multiple radio modules under identical environmental conditions. For a comprehensive treatment of ANOVA methodology, we refer the reader to \cite{montgomery2022design}. Environment-level F-statistics as shown in Table~\ref{tab:f_statistics_comparison} are considerable (BLE: 149,256; CC1200: 413,499; CC2538: 4,489,917; nRF52840: 147,414), indicating significant differences in RSSI across environments ($p < 0.05$). A p-value below 0.05 is commonly used to denote statistical significance, meaning there is less than a 5\% probability that the observed differences occurred by chance. Therefore, these results are statistically significant. The combination of high F-statistics and low p-values confirms that environmental factors strongly influence RSSI, rather than the observed differences being random fluctuations. Hardware-level F-statistics are smaller but still significant (BLE: 1,542.3; CC1200: 3,218.7; CC2538: 12,367.9; nRF52840: 1,434.8) demonstrating high measurement repeatability and limited hardware variability. These findings indicate that while environmental conditions are the primary driver of RSSI variations, hardware differences contribute a smaller, measurable effect.

\begin{table}[h!]
\centering
\caption{F-statistics for RSSI means and radio consistency. $df_{in}$ denotes within-group degrees of freedom; the between-group degrees of freedom is 4, with a p-value of $1.976 \times 10^{-32}$.}

\begin{tabular}{|l|l|l|r|}
\hline
\textbf{Radio} & \textbf{Analysis Level} & \textbf{F-Statistic} & \textbf{df$_{in}$} \\
\hline
BLE        & Env.   & $149{,}256 \pm 105{,}000$   & 89,876 \\
BLE        & Device & $1{,}542 \pm 1{,}090$       & 89,876 \\
CC1200     & Env.   & $413{,}499 \pm 292{,}000$   & 89,166 \\
CC1200     & Device & $3{,}219 \pm 2{,}280$       & 89,166 \\
CC2538     & Env.   & $4{,}489{,}917 \pm 3{,}180{,}000$ & 812,511 \\
CC2538     & Device & $12{,}368 \pm 8{,}750$      & 812,511 \\
nRF52840   & Env.   & $147{,}414 \pm 104{,}000$   & 691,018 \\
nRF52840   & Device & $1{,}435 \pm 1{,}010$       & 691,018 \\
\hline
\end{tabular}
\label{tab:f_statistics_comparison}
\end{table}

Accordingly, environment effects dominate across all radio platforms. This is evident from the environmental effect sizes in Table~\ref{tab:radio_factors}: CC2538 ($\eta^2_{\text{env}} = 0.120$, $\omega^2_{\text{env}} = 0.118$), CC1200 ($0.210$, $0.208$), BLE ($0.090$, $0.088$), and nRF52840 ($0.300$, $0.295$), all exceeding their corresponding node effects. Node contributions remain lower for all radios, ranging from 0.020 (BLE) to 0.180 (CC1200) in $\eta^2_{\text{node}}$. Consistently, all platforms are classified as environment-dominant in the table.

\begin{table}[h!]
\centering
\caption{Comparison of effect sizes ($\eta^2$ and $\omega^2$) for environment and node factors, indicating that environments are dominant}
\begin{tabular}{|l|c|c|c|c|c|}
\hline
Radio & $\eta^2_{\text{env}}$ & $\omega^2_{\text{env}}$ & $\eta^2_{\text{node}}$ & $\omega^2_{\text{node}}$\\
\hline
CC2538  & 0.120 & 0.118 & 0.070 & 0.068 \\
CC1200  & 0.210 & 0.208 & 0.180 & 0.176  \\
BLE     & 0.090 & 0.088 & 0.020 & 0.019  \\
nRF52840 & 0.300 & 0.295 & 0.130 & 0.127 \\
\hline
\end{tabular}
\label{tab:radio_factors}
\end{table}

Both $\eta^2$ and $\omega^2$ are effect size measures used in ANOVA. $\eta^2$ estimates the proportion of variance explained but tends to be slightly biased upward, especially in small samples. $\omega^2$ applies a correction for this bias and is therefore more conservative and closer to the true population effect size.

A Tukey Honestly Significant Difference (HSD)  post-hoc test was performed after ANOVA to identify pairwise group differences across radios. Nearly all comparisons were significant (39/40), indicating strong differences between groups. All comparisons were significant ($p=0$) for CC2538, BLE, and nRF52840, while CC1200 had one non-significant pair (LK vs RV, $p = 0.197$). Overall, results show a strong and consistent effect of the grouping factor across all radios, see Table \ref{tab:post-hoc}.

\begin{table}[h]
\centering
\caption{Summerized Tukey HSD post-hoc results across radio types}
\begin{tabular}{|l|c|c|c|}
\hline
\textbf{Radio} & \textbf{Total Pairs} & \textbf{Significant} & \textbf{Not Significant} \\
\hline
CC2538   & 10 & 10 & 0 \\
CC1200   & 10 & 9  & 1 \\
BLE      & 10 & 10 & 0 \\
nRF52840 & 10 & 10 & 0 \\
\hline
\end{tabular}
\label{tab:post-hoc}
\end{table}

The frequency of exceeding threshold values in the deployment environments suggest that sustained link instability rather than isolated measurement noise are the reason for the strong link quality fluctuations. In contrast, the garden environment demonstrates relatively stable RSSI dynamics with low RSSI variation. Near line-of-sight conditions and moderate obstruction allow the EMA to closely track the raw RSSI signal, resulting in a significantly lower outlier rate (\(p < 0.01\)). This characteristic highlights the impact of environmental openness on signal stability. Lake and river environments exhibit intermediate RSSI dynamics. While open water surfaces generally favor stable propagation, intermittent reflections and surface-induced scattering produce sporadic RSSI deviations. These deviations appear as isolated outliers rather than continuous instability, which is reflected in moderate outlier rates of the detection.

Across all environments, radio-specific characteristics further exacerbate RSSI dynamics. CC1200 radios exhibit smoother RSSI trajectories and fewer extreme deviations, suggesting higher robustness to environmental variability. Conversely, nRF52840 and BLE radios show increased sensitivity to multipath environments, resulting in higher outlier rates. These observations demonstrate that RSSI dynamics and outlier patterns are jointly influenced by environmental conditions and radio hardware characteristics.

\subsection{Comparison of Radio Outlier Rates}

To assess the outlier rates for each radio and deployment environment, we set the confidence level to $\delta = 0.05$. This means that at a confidence coefficient of 0.95, the fluctuation in link quality modeled by the random variable $\mathbf{z}_t$ is considered normal. As mentioned at the end of the previous section, a larger value of $k$ indicates higher sensitivity (a link that is prone to outliers).

The Outlier Rate comparison highlights distinct differences in performance consistency among the radios. The CC2538 radios exhibit the lowest outlier rates across all environments, often near zero (as low as 0.002), indicating relatively stable and reliable operation. BLE radios show moderate outlier rates, with the highest observed in the FR deployments (0.016), suggesting occasional instability under certain conditions. The CC1200 radios have slightly higher and more consistent outlier rates compared to CC2538, typically ranging up to 0.017, although the rates remain generally low. The nRF52840 radios show the highest outlier rates, reaching up to 0.026 in certain environments. Considering $\alpha$ and $k$ alongside outlier rates, it is evident that radios with high $\alpha$ and low $k$, such as nRF52840, do not necessarily produce fewer outliers, while radios with lower $\alpha$ and higher $k$, such as CC2538, often experience fewer deviations. Overall, CC2538 provides the most consistent performance, BLE and CC1200 are moderately reliable, and nRF52840 is comparatively less stable in terms of outlier occurrences. This is summarized in Table~\ref{tab:outlier_stability}.

\begin{table}[!h]
\caption{Outlier rate rates for all radios and deployment environments in terms of $\alpha$ and $k$.}
\label{tab:outlier_stability}
\centering
\begin{tabular}{llcccccc}
\toprule
Radio & Env &  Outlier Rate & $\alpha$ & $k$ \\
\midrule
BLE       & BG  & 0.007 & 0.775 & 3.367 \\
BLE       & FR & 0.016 & 0.834 & 3.304 \\
BLE       & GG & 0.014 & 0.934 & 3.220 \\
BLE       & PP & 0.004 & 0.624 & 3.522 \\
BLE       & RV & 0.011 & 0.770 & 3.372 \\

CC1200    & BG & 0.012 & 0.950 & 3.201 \\
CC1200    & FR & 0.017 & 0.950 & 3.192 \\
CC1200    & LK & 0.007 & 0.950 & 3.192 \\
CC1200    & RA & 0.012 & 0.950 & 3.192 \\
CC1200    & RV & 0.014 & 0.950 & 3.200 \\

CC2538    & BG & 0.005 & 0.599 & 3.516 \\
CC2538    & FR & 0.008 & 0.601 & 3.498 \\
CC2538    & LK & 0.011 & 0.597 & 3.510 \\
CC2538    & RA & 0.002 & 0.594 & 3.552 \\
CC2538    & RV & 0.013 & 0.602 & 3.499 \\

nRF52840  & BG & 0.019 & 0.950 & 3.191 \\
nRF52840  & FR & 0.022 & 0.950 & 3.194 \\
nRF52840  & GG & 0.024 & 0.950 & 3.194 \\
nRF52840  & LK & 0.026 & 0.950 & 3.191 \\
nRF52840  & CA & 0.009 & 0.950 & 3.189 \\
\bottomrule
\end{tabular}
\end{table}

\subsection{Comparison with Baselines}

\begin{figure}[!h]
\centering
\includegraphics[width=0.95\linewidth]{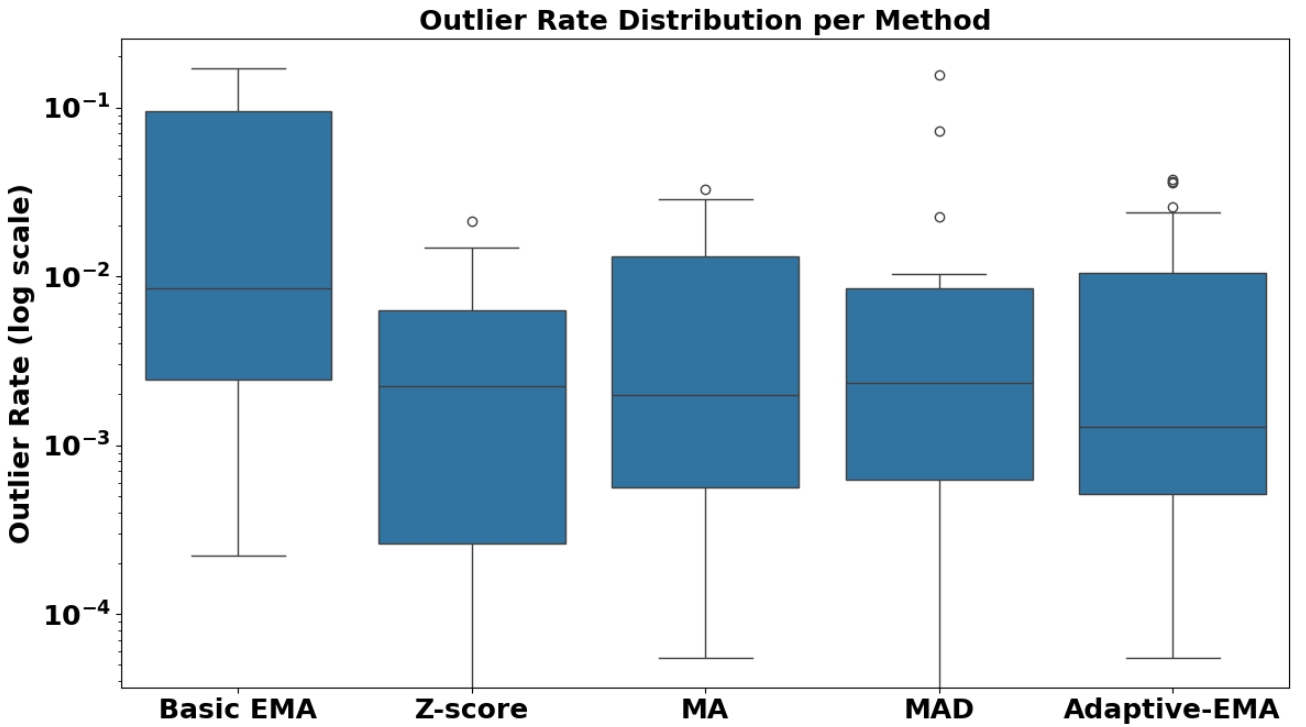}
\caption{Distribution of outlier detection rates for five anomaly detection methods across four radio platforms and fifteen environmental scenarios.}
\label{fig:outlier_boxplot}
\end{figure}

In this section we compare the performance of our approach (Adaptive EMA) based on relative behavior of variation in the signal with four baseline outlier detection methods: Basic EMA (our approach with a fixed $\alpha$ as a threshold), interval estimation based on z-score, Moving Average, Median Absolute Deviation (MAD), and Adaptive EMA (our approach). Figure~\ref{fig:outlier_boxplot} displays the distribution of outlier detection rates aggregated for the four radios and the fifteen deployment environments. The vertical axis is represented on a logarithmic scale to allow visualization of detection rates spanning several orders of magnitude. In each boxplot, the central horizontal line indicates the median value, the box boundaries represent the first and third quartiles (Q1 and Q3), and the whiskers extend to the typical minimum and maximum values, while individual points denote extreme observations. 

Basic EMA exhibits the highest median detection rate, approximately \(10^{-2}\), with a wide interquartile range and whiskers extending close to \(10^{-1}\). This indicates high sensitivity but also substantial variability, suggesting potential over-detection or false positives. In contrast, z-score---$\mathbf{z}_t = ( \mathbf{R}_t-\eta_x)/\sigma_x)$---shows the lowest median and narrowest spread near \(10^{-3}\), reflecting a conservative detection minimizing false positives but risking missing subtle anomalies. Moving Average lies between z-score and Basic EMA, with moderate median values and a wider spread than z-score, capturing more anomalies while maintaining moderate stability. MAD has a similar median to MA but features several extreme outliers, indicating occasional high detection rates and higher variability in some scenarios. Finally, Adaptive EMA  (outlier threshold = $k\sigma{_z}$) maintains a median detection rate comparable to MA and MAD but with a narrower interquartile range and fewer extreme values. This demonstrates a balanced trade-off between sensitivity and stability, enabled by its dynamic smoothing adjustment. Overall, Adaptive EMA demonstrates more consistent anomaly-detection behavior across varying radios and environments and shows comparatively stronger robustness and statistical stability.
.

\section{Conclusion}
\label{sec:conclude}

This paper presented a comprehensive experimental analysis of RSSI fluctuations and statistical outlier behavior across heterogeneous IoT radio platforms deployed in diverse outdoor environments. By integrating a mathematical RSSI abstraction with exponential moving average (EMA) smoothing and statistically grounded outlier detection, the study examined how environmental complexity and radio hardware characteristics jointly influence signal stability. Experimental results showed that environments with high structural and propagation complexity exhibit higher RSSI variance and increased outlier occurrence compared to more open environments such as gardens. The proposed outlier detection  provides a compact and interpretable representation of RSSI anomalies, enabling systematic comparison across radio platforms and deployment scenarios. Among the evaluated radios, the CC2538 consistently demonstrates the greatest robustness to environmental variability in terms of lowest observed outlier rates. The nRF52840 also exhibits high overall stability but shows increased sensitivity in multipath-rich environments, with higher outlier rates, while BLE and CC1200 radios display moderate and more consistent performance. These findings provide practical insights for IoT and sensor network designers, supporting informed radio selection, environment-aware deployment planning, and adaptive communication strategies. The lightweight nature of the proposed framework makes it well suited for resource-constrained IoT devices and long-term outdoor monitoring applications, where predictable link behavior is essential.

Future work will explore adaptive thresholding techniques, multi-metric link-quality fusion, and machine learning–assisted anomaly classification to further enhance robustness under highly dynamic propagation conditions.

\balance

\bibliographystyle{IEEEtran}
\bibliography{references}

\end{document}